\documentclass{PoS}

\title{HESS J1641-463, a very hard spectrum TeV gamma-ray source in the Galactic plane.}

\ShortTitle{HESS J1641-463}

\author{\speaker{I. Oya}\\
        DESY, Zeuthen, Germany\\
        E-mail: \email{igor.oya.vallejo@desy.de}}
\author{S. Casanova \\
	Instytut Fizyki J\c{a}drowej PAN, Krak{\'o}w, Poland \\ 
	Max-Planck-Institut f\"ur Kernphysik, Heidelberg, Germany\\
	E-mail: \email{sabrina.casanova@ifj.edu.pl}, \email{sabrina.casanova@mpi-hd.mpg.de}}
\author{F. Aharonian\\
	 Max-Planck-Institut f\"ur Kernphysik, Heidelberg, Germany\\ 
	 Dublin Institute for Advanced Studies, Dublin 2, Ireland}
	\author{M. Dalton\\
	Universit\'e Bordeaux 1, CNRS/IN2P3, Gradignan, France\\
	Now at: Active Space Technologies GmbH, Berlin, Germany }
\author{For the H.E.S.S. collaboration}

\newcommand{\ensuretext[1]}{\ensuremath{\text{#1}}}

\newcommand{\g}{\ensuremath{\gamma}}%

\newcommand{\hess}{\textsc{H.E.S.S.}}
\newcommand{\hessjlong}{HESS\,J1641$-$463}
\newcommand{\hessj}{J1641$-$463}
\newcommand{\fourty}{HESS\,J1640$-$465}

\newcommand{\fluxHESSaatOneTeV}{\ensuremath{\phi_0 =  (3.91 \pm 0.69_{\rm stat} \pm 0.78_{\rm sys}) \times 10^{-13} \rm cm^{-2}s^{-1}TeV^{-1}}}
\newcommand{\intfluxHESSaoneTeV}{\ensuremath{\phi\,(\rm E > 1\, TeV) = (3.64 \pm 0.44_{\rm stat} \pm 0.73_{\rm sys}) \times 10^{-13} \rm cm^{-2}s^{-1}}}

\newcommand{\GammaHESSa}{\ensuremath{\Gamma = 2.07 \pm 0.11_{\rm stat} \pm 0.20_{\rm sys}}}

\abstract{HESS J1641$-$463 is a unique source discovered by the High Energy Stereoscopic System (H.E.S.S.) telescope array in the multi-TeV domain. The source had been previously hidden in the extended tail of emission from the bright nearby source HESS J1640$-$465. However, the analysis of the very-high-energy (VHE) data from the region at energies above 4 TeV revealed this new source at a significance level of 8.5$\sigma$. HESS J1641$-$463 showed a moderate flux level F(E $>$ 1 TeV) = (3.64 $\pm$ 0.44$_{stat}$ $\pm$ 0.73$_{sys}$) $\times$ 10$^{-13}$ cm$^{-2}$s$^{-1}$, corresponding to 1.8\% of the Crab Nebula flux above the same energy, and a hard spectrum with a photon index $\Gamma$ = 2.07 $\pm$ 0.11$_{stat}$ $\pm$ 0.20$_{sys}$. The light curve was investigated for evidence of variability, but none was found on both short (28-min observation) and long (yearly) timescales. HESS J1641-463 is positionally coincident with the radio supernova remnant (SNR) G338.5$+$0.1. There is no clear X-ray counterpart of the SNR, although {\it Chandra} and XMM-{\it Newton} data reveal some weak emission that may be associated.  If the emission from HESS J1641$-$463 is produced by cosmic ray protons colliding with the ambient gas, then the proton spectrum extends up to  0.1 PeV (99\% confidence level) and likely to higher energies, > 0.27 PeV (90 \% confidence level). If this is the case, then HESS J1641-463 may be a member of a larger source population contributing to the Galactic cosmic-ray flux around the knee.}

\FullConference{The 34th International Cosmic Ray Conference,\\
		30 July- 6 August, 2015\\
		The Hague, The Netherlands}

\begin{document}

\section{H.E.S.S. observations and analysis}

The data used for the analysis described here were taken with  High Energy Stereoscopic System (H.E.S.S.) between 2004 and 2011, amounting to an acceptance-corrected livetime of 72 hours. The events were reconstructed using a Hillas parameter technique \cite{hillas}. The results where crosschecked using independent methods \cite{2009APh....31..383O, 2009APh....32..231D}, yielding consistent results.

\begin{figure}
\label{maps}
\centering
\begin{minipage}{0.63\textwidth}
\centering
\includegraphics[width=\textwidth]{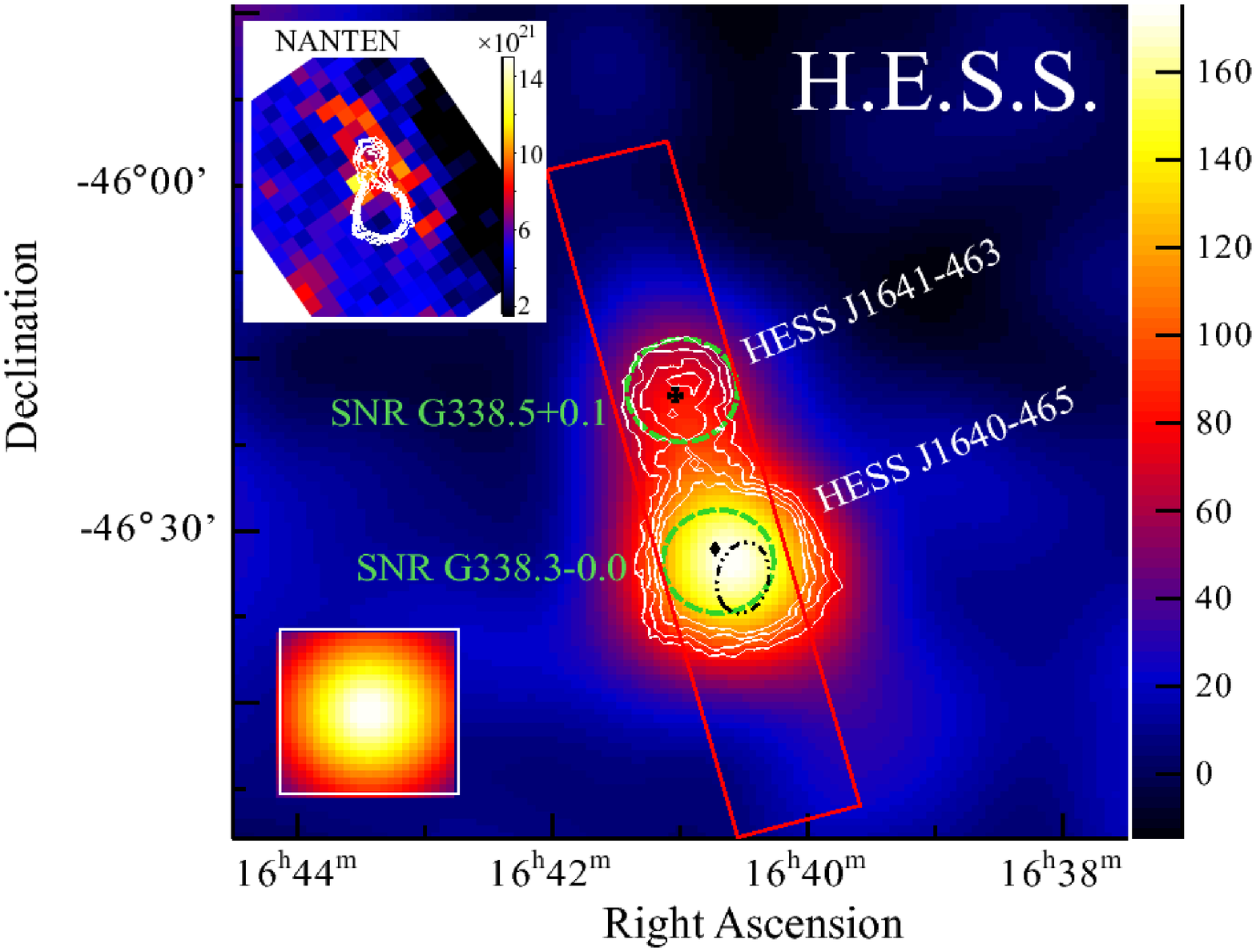}
\end{minipage}\hfill
\begin{minipage}{0.30\textwidth}
\centering
\includegraphics[width=\textwidth]{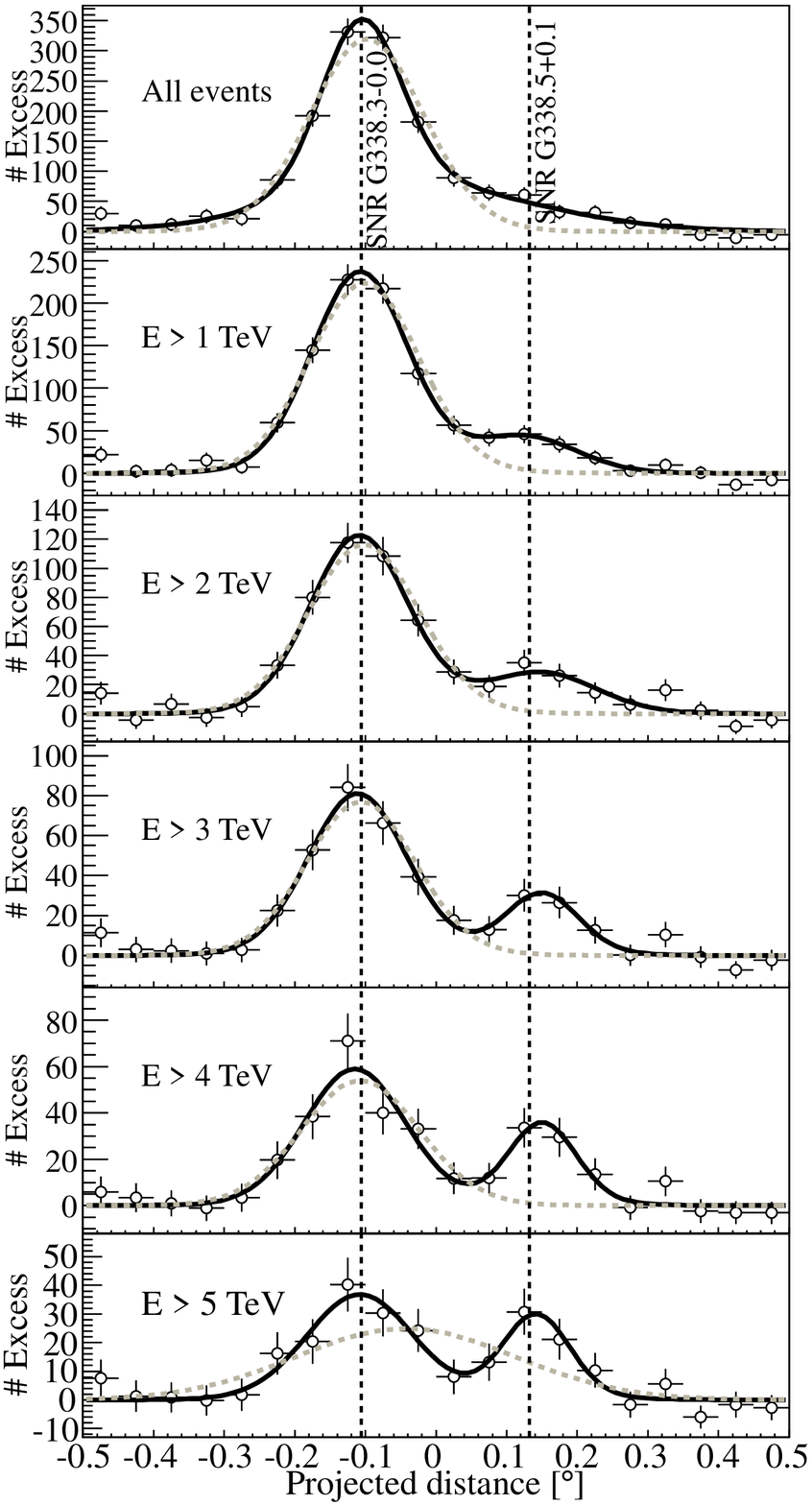}
\end{minipage}
\caption{{\it Left:} Map of excess events with energies E\,$>$\,4\,TeV for the
    region around \hessj\ smoothed with the 
    instrument PSF (shown inside the
    white box). The white contours indicate the significance of
    the emission at the 5, 6, 7 and $8\sigma$ level. The black cross
    indicates the best fit position of
    the source, the green dashed circles show the positions and
    approximate extensions of the two nearby SNRs, the black diamond
    the position of PSR J1640$-$4631, the dash-dotted black ellipse the
    95\% confidence error position of 1FHL J1640.5$-$4634, and the red
    box indicates the area for the extraction of the profiles shown in the right panel. The upper left inset shows a map of the distribution of
    the column density of molecular hydrogen in units of cm$^{-2}$,
    estimated from the NANTEN CO(1$-$0) data, together with the
    \hess\ significance contours. {\it Right:} Distribution of VHE $\g$-ray excess profiles and Gaussian
    fits for the red rectangular
    slice shown in the left panel. Vertical lines show the
    position of the SNR 338.3$-$0.0 and G338.5+0.1. Figures from \cite{paper1641}.}
\end{figure}

\hessjlong (\hessj\ hereafter) was not visible in the
original \fourty\ images, without energy
cut in the events, due to the much brighter, nearby source \fourty. However, thanks to the
improved \hess\ point spread function (PSF) at higher energies, and to
its hard spectrum, \hessj\ was clearly visible when increasing the energy threshold, where the contamination from \fourty\ was reduced. The VHE \g-ray excess image
obtained for E $>$ 4 TeV is shown in Fig.\ \ref{maps}, left, where the
background level is estimated following the ring background model
\cite{2007A&A...466.1219B}. Figure~\ref{maps}, right, shows the projection of the excess
events in the rectangular region shown in Fig. \ref{maps}, left, for
different energy bands.

We fitted the differential VHE \g-ray spectrum of \hessj\ with a power-law function  ${\rm d}N/{\rm d}E = \phi_0 \times (E/1
\,\rm TeV)^{-\Gamma}$ in the range from 0.64 TeV to 100 TeV
using the forward-folding technique \cite{2001A&A...374..895P}, the best fitting parameters being \fluxHESSaatOneTeV\ and \GammaHESSa.  The flux level is
\intfluxHESSaoneTeV, corresponding to 1.8\% of the Crab Nebula flux
above the same energy. The light curve was investigated for evidence of variability, but none was found on both short (28-min observation) and long (yearly) timescales.

\section{Search for multiwavelength counterpart}

\subsection{In the radio band}
\hessj\ is found within the bounds of the supernova remnant (SNR) G338.5+0.1
\cite{2009BASI...37...45G}. No precise estimates on the age of this SNR exist. Depending on its physical size, however, its age could range between 1.1 to 1.7 kyr \cite{paper1641}.

The distribution of molecular gas around \hessj\ is shown
in the top left inset of Fig.~\ref{maps}.  This distribution is
obtained by integrating the CO 1$\rightarrow$0 rotational line
emission over a range in velocity between $-$40
km/s to $-$30 km/s
\cite{2001PASJ...53.1003M,2004ASPC..317...59M}. The choice of this
range is motivated by the presence of dense molecular cloud (MC) clumps in
the region \cite{2012AIPC.1505..277D}. Using the
model for the Galactic rotation curves by Kothes \& Dougherty \cite{2007A&A...468..993K}
the gas is located at a distance of $\sim$ 11 kpc. Assuming a ratio $X_{CO->N_{H_2}}=1.5 \times {10}^{20}$ between the CO
velocity integrated intensity and the column density of molecular gas,
$N_{H_2}$, the total column density from the extraction region of
\hessj\ is $ 1.7 \times {10}^{22}$ cm$^{-2}$. At 11 kpc the density
and the total mass of the gas clumps inside the source region are about 100 cm$^{-3}$ and $ 2.4 \times {10}^5$
solar masses, respectively.

\subsection{In the X-rays}
No candidate for an X-ray counterpart of \hessj\ was found in existing
catalogs. Two data sets from {\it Chandra} (ObsID 11008 and ObsID 12508)  and one from
XMM-\textit{Newton} (ObsID 0302560201) were inspected in order to search for an
X-ray counterpart of \hessj. The outcome of this analysis, described in detail in ref. \cite{paper1641}, is that 
no obvious counterpart of \hessj\  could be established neither in the {\it Chandra} nor the XMM-\textit{Newton} data.

\subsection{In the high energy \g-rays}

Lemoine-Goumard et~al. \cite{1641Fermi}  report the detection of two distinct sources with Fermi-LAT above 100 MeV, corresponding to the positions of \hessj\ and \fourty. The softest emission in this region as seen by Fermi-LAT corresponds to \hessj, which is well fitted with a power law of index $\Gamma$ = 2.47 $\pm$ 0.05 $\pm$ 0.06 and presents no significant gamma-ray signal above 10 GeV. The connection between
this  steep  spectrum  and  the hard
H.E.S.S.  spectrum  remains  unclear, although the ${\it{Fermi}}$-LAT and H.E.S.S results suggest two
different  mechanisms, or  sources,  producing  the
\g-ray  emission.

\section{Scenarios to explain the \hessj\ emission}

\begin{figure}
  \centering
  \includegraphics[width=0.65\textwidth]{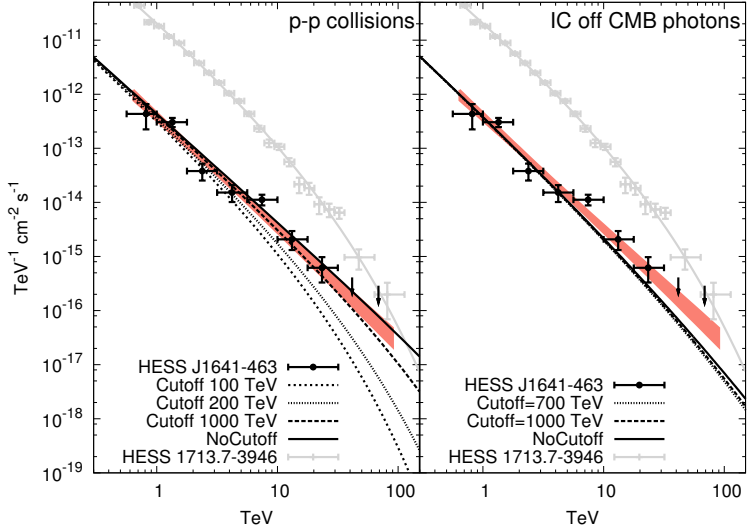}
  \caption{Differential \g-ray spectrum of \hessj\ together with
    the expected emission from p-p collisions (left) and IC off CMB
    photons (right). The pink area represents the $1\sigma$ confidence
    region for the fit to a power law model, the black data points the
    H.E.S.S. measured photon flux, the
    arrows the 95\% CL upper limits on the flux level, and the black
    curves the expected emission from the models. For comparison, the gray data
    points and curve represent the archival data of SNR RX
    J1713.7$-$3946 \cite{2007A&A...464..235A}. Figure from \cite{paper1641}.}
  \label{EmissionM}
\end{figure}

A first possible scenario to explain the \hessj\ emission is that the SNR\,G338.5+0.1  is young and is able to accelerate particles up to hundreds of TeV. The left panel of Fig.~\ref{EmissionM} shows the
comparison between the \hess\ spectrum and the spectrum produced by
accelerated protons from G338.5+0.1, interacting with the ambient
gas. The predicted spectra are calculated using the parametrization of Kelner et~al.
\cite{2006PhRvD..74c4018K}, assuming a proton spectrum with a
power-law slope of $-$2.1. The 99\% confidence level
(CL) lower limit on the proton cutoff energy corresponds to 100 TeV. Remarkably, the \g-ray
spectrum of \hessj\ is harder than that observed from the young SNR
RXJ1713$-$4936 at energies above few TeV
\cite{2007A&A...464..235A}. With a proton spectrum extending almost up to 1 PeV, \hessj\ may be a representative of a new source type that is
contributing significantly to the galactic cosmic-rays flux around the
knee.

%Remarkably, if the ambient density
%is $n = 1~\mathrm{cm}^{-3}$ and between 10 to 30 \% of the energy of
%the SN explosion is converted to CRs, then this SNR would explain the
%Galactic CR luminosity above a few GeV.

If G338.5+0.1 is older, VHE
protons accelerated by the young SNR\,G338.3$-$0.0, positionally
coincident with \fourty\, \cite{2014MNRAS.439.2828A}, could have
already reached the dense MC coincident with
\hessj. This would explain the relatively high brightness of
\hessj\ in comparison with \fourty\ at high energies. In
such a scenario, the much younger adjacent
G338.3$-$0.0 would be a major source of CRs.

Electrons of hundreds of TeV inverse Compton (IC) scattering off
the cosmic microwave background photons (CMB) could explain the
emission from \hessj. These electrons would be accelerated either in
G338.5+0.1 or in the PWN associated to the young energetic
pulsar, PSR J1640$-$4631 \cite{2014ApJ...788..155G}. Even assuming a
pure power law for the primary electron spectrum, the cross section
for IC scattering decreases at high energies resulting in a break in
the \g-ray spectrum at multi TeV energies, not
observed in the spectrum of \hessj. The
predicted IC radiation, shown in the right panel of
Fig. \ref{EmissionM}, was obtained by assuming that the electron
cooled spectrum is a power law of spectral index $-$3.14. The 99\% CL lower limit on the cutoff
energy corresponds to 700 TeV. We note that it is extremely
difficult to accelerate electrons in SNRs to such energies as hundred TeV
electrons suffer severe synchrotron losses in the amplified magnetic
fields of the acceleration sites.

\section{Conclusions}
We present a new unique VHE
source, showing one of the hardest \g-ray spectra ever found at these
energies. In order to
explain the observed VHE \g-ray spectrum, scenarios where protons are
accelerated up to hundreds of TeV at either G338.5+0.1 or
G338.3$-$0.0, and then interact with local gas or nearby massive
MCs are the most compelling ones. In such scenario, \hessj\ would be a representative of a source class contributing to the Galactic cosmic-ray flux around the knee. Otherwise, a leptonic emission-based scenario cannot be formally ruled out, although it is severely constrained by the absence of breaks in the \g-ray spectrum and by the high efficiency required for an accelerator located within the SNR, where strong losses through synchrotron emission are expected. 

\section*{Acknowledgements}
  The support of the Namibian authorities and of the University of
  Namibia in facilitating the construction and operation of
  H.E.S.S. is gratefully acknowledged, as is the support by the German
  Ministry for Education and Research (BMBF), the Max Planck Society,
  the German Research Foundation (DFG), the French Ministry for
  Research, the CNRS-IN2P3, and the Astroparticle Interdisciplinary
  Programme of the CNRS, the U.K. Science and Technology Facilities
  Council (STFC), the IPNP of the Charles University, the Czech
  Science Foundation, the Polish Ministry of Science and Higher
  Education, the South African Department of Science and Technology
  and National Research Foundation, and by the University of
  Namibia. We appreciate the excellent work of the technical support
  staff in Berlin, Durham, Hamburg, Heidelberg, Palaiseau, Paris,
  Saclay, and in Namibia in the construction and operation of the
  equipment. Sabrina Casanova acknowledges the support from the Polish National Science Center under the Opus Grant \mbox{UMO-2014/13/B/ST9/00945}


\begin{thebibliography}{99}

\bibitem{hillas}{Hillas}, A.~M. 1995, Proc. of the 19th ICRC (La Jolla), 3, 445

\bibitem{2009APh....31..383O}
{Ohm}, S., {van Eldik}, C., \& {Egberts}, K. 2009, Astroparticle Physics, 31,
  383

\bibitem{2009APh....32..231D}
{de Naurois}, M., \& {Rolland}, L. 2009, Astroparticle Physics, 32, 231

\bibitem{2007A&A...466.1219B}
{Berge}, D., {Funk}, S., \& {Hinton}, J. 2007, A\&A, 466, 1219

\bibitem{2001A&A...374..895P}
{Piron}, F., {Djannati-Atai}, A., {Punch}, M., {et~al.} 2001, A\&A, 374, 895

\bibitem{2009BASI...37...45G}
{Green}, D.~A. 2009, Bulletin of the Astronomical Society of India, 37, 45

\bibitem{paper1641}
{Abramowski}, A.,  {et~al.} 2014, ApJL, 794, L1


\bibitem{2001PASJ...53.1003M}
{Matsunaga}, K., {et~al.} 2001, PASJ, 53, 1003

\bibitem{2004ASPC..317...59M}
{Mizuno}, A., \& {Fukui}, Y. 2004, in Astronomical Society of the Pacific
  Conference Series, Vol. 317, 59




\bibitem{2012AIPC.1505..277D}
{de Wilt}, P., {et~al.} 2012, in American Institute
  of Physics Conference Series, Vol. 1505, American Institute of Physics
  Conference Series, 277--280


\bibitem{2007A&A...468..993K}
{Kothes}, R., \& {Dougherty}, S.~M. 2007, A\&A, 468, 993


\bibitem{1641Fermi}
{Lemoine-Goumard, M.}, {et~al.} 2014, ApJL, 794, L16



\bibitem{2007A&A...464..235A}
{Aharonian}, F. A.,  {et~al.} 2007, A\&A, 464, 235


\bibitem{2006PhRvD..74c4018K}
{Kelner}, S.~R., {Aharonian}, F.~A., \& {Bugayov}, V.~V. 2006, PhRvD, 74, 034018


\bibitem{2014MNRAS.439.2828A}
{Abramowski}, A.,  {et~al.} 2014, MNRAS,
  439, 2828


\bibitem{2014ApJ...788..155G}
{Gotthelf}, E.~V., {et~al.} 2014, ApJ,
  788, 155




\end{thebibliography}
\end{document}